\documentclass[twocolumn]{aastex63}

\providecommand{\sorthelp}[1]{}
\submitjournal{ApJ}
\accepted{October 14, 2022}

\shorttitle{PAH Emission in $\lambda$-Orionis}
\shortauthors{Chuss et al.}
\graphicspath{{./}{figures/}}

\begin{document}

\title{Tracing PAH Emission in $\lambda$-Orionis Using COBE/DIRBE Data}

\correspondingauthor{David T. Chuss}
\email{david.chuss@villanova.edu}

\author[0000-0003-0016-0533]{David T. Chuss}
\affil{Department of Physics, Villanova University, 800 E. Lancaster Ave., Villanova, PA 19085, USA}

\author[0000-0001-7449-4638]{Brandon S. Hensley}
\affil{Department of Astrophysical Sciences,  Princeton University, Princeton, NJ 08544, USA}

\author[0000-0001-9835-2351]{Alan J. Kogut}
\affil{Observational Cosmology Laboratory, Code 665, NASA Goddard Space Flight Center, Greenbelt, MD 20771}

\author[0000-0001-8819-9648]{Jordan A. Guerra}
\affil{Department of Physics, Villanova University, 800 E. Lancaster Ave., Villanova, PA 19085, USA}

\author[0000-0001-9694-1718]{Hayley C. Nofi}
\affiliation{Department of Astrophysics and Planetary Science, Villanova University \\
800 Lancaster Avenue,
Villanova, PA, 19085, USA}

\author[0000-0001-5389-5635]{Javad Siah}
\affil{Department of Physics, Villanova University, 800 E. Lancaster Ave., Villanova, PA 19085, USA}

\begin{abstract}
We use archival COBE/DIRBE data to construct a map of polycyclic aromatic hydrocarbon (PAH) emission in the $\lambda$-Orionis region.  The presence of the 3.3~\micron\ PAH feature within the DIRBE 3.5~\micron\ band and the corresponding lack of significant PAH spectral features in the adjacent DIRBE bands (1.25, 2.2, and 4.9~\micron) enable estimation of the PAH contribution to the 3.5~\micron\ data. Having the shortest wavelength of known PAH features, the 3.3~\micron\ feature probes the smallest PAHs, which are also the leading candidates for carriers of anomalous microwave emission (AME). We use this map to investigate the association between the AME and the emission from PAH molecules. 
We find that the spatial correlation in $\lambda$-Orionis is higher between AME and far-infrared dust emission (as represented by the DIRBE 240\,\micron\ map) than it is between our PAH map and AME. This finding, in agreement with previous studies using PAH features at longer wavelengths, is in tension with the hypothesis that AME is due to spinning PAHs. However, the expected correlation between mid-infrared and microwave emission could potentially be degraded by different sensitivities of each emission mechanism to local environmental conditions even if PAHs are the carriers of both.
\end{abstract}


\section{Introduction} \label{sec:intro}
Anomalous microwave emission (AME) refers to the $\sim$10-60\,GHz radiation measured throughout the Galaxy in excess of the well-characterized synchrotron and thermal bremsstrahlung that otherwise dominate the emission from the interstellar medium in this frequency range \citep{Dickinson2018}.  Peaking between 15 and 40\,GHz, AME was initially observed in experiments that measured the cosmic microwave background (CMB) and found to be spatially correlated with far-infrared (FIR) thermal emission from dust \citep{Kogut1996,deOliveira1997,Leitch1997}. Since then, increasingly sensitive CMB measurements have demonstrated that AME is ubiquitous in interstellar medium \citep{Bennett2003}, and an all-sky map of the distribution of AME has been produced with Planck data \citep{PlanckintXII2014}. More recently, multiwavelength data in the region of $\lambda$-Orionis has been utilized to create a map of AME in this region with improved component separation \citep{Bell2019PASJ,Cepeda2020}. 

Though the frequency spectrum of AME is increasingly well characterized, its origin is still not fully understood. The leading physical explanation for AME is the ``spinning dust'' model \citep{Draine1998b}. In this scenario, small, rapidly spinning dust grains emit electric dipole radiation \citep{Draine1998a, Ali-Haimoud2009, Ysard2010, Hoang2011}. Alternate explanations include magnetic dipole emission from fluctuations in iron inclusions in larger dust grains \citep{Draine1997, Draine2013}, emission from two-level systems in dust \citep{Jones2009}, and magnetic dipole emission from magnetic nanoparticles \citep{Draine2013}.

If the spinning dust paradigm is correct, a prime candidate for the carriers of the emission is the class of small dust grains (or large molecules) known as ``polycyclic aromatic hydrocarbons'' \cite[PAHs;][]{Tielens2008}. PAHs consist of tiled carbon rings arranged in a planar structure with hydrogen atoms bonded at the perimeter.  These particles have $\lesssim 1000$ atoms and typically are less than a few nanometers in size.  Such small particles are capable of spinning at $\sim$GHz frequencies given excitation conditions in the interstellar medium \citep{Draine1998a}.

PAHs are identifiable via a series of broad emission features in the mid-infrared region of the spectrum \citep{Tielens2008}. These features correspond to bending and stretching of the lattice and C-H bond of the PAHs.  Most prominent of these include features at 3.3 \micron, due to the C-H stretching modes; 6.2 and 7.7 \micron\, due to vibrations of the carbon sheet; 8.6 \micron\ from in-plane C-H bending; and 11.3, 12.0, 12.7, and 13.6 \micron\ from out-of-plane C-H bending \citep{Draine2011}.

One method for testing the hypothesis that PAHs produce the AME is to compare the spatial distribution of AME to tracers of PAHs. \citet{Hensley2016} tested the correlation between PAH emission in the Wide-field Infrared Survey Explorer (WISE) W3 band using an all-sky map of diffuse emission created from the WISE data set \citep{Meisner2015}. This band has significant contribution from the 11.3 \micron\ PAH feature. \citet{Hensley2016} found that the AME is more tightly correlated with the FIR dust continuum than with the 12 \micron\ emission.

A potential explanation for this finding within the spinning dust paradigm is that the AME arises from spinning grains that are not PAHs. It has been shown that nano-silicate grains could exist in sufficient abundance to explain the AME \citep{Hoang2016,Hensley2017}, although \citet{Ysard2022} have recently argued that these grains cannot reproduce the AME spectral energy distribution in detail. Spinning grains with a magnetic dipole moment could also contribute to the observed AME, though they are unable to account for the entirety of the emission \citep{Hoang2016b,Hensley2017}. Another possibility is that PAH emission physics and AME depend differently on local interstellar conditions, eroding the correlation between the two even if they originate from the same grains \citep{Hensley2022, Ysard2022}.

One future area of progress is improving the precision of the AME maps. For example, \citet{Cepeda2020} have examined the AME spectrum in $\lambda$-Orionis using additional data sets to fit the spectral energy distribution of the region across the microwave part of the spectrum. A second area is more sophisticated fitting techniques. \citet{Bell2019PASJ} utilize hierarchical Bayesian inference to fit spectral energy distributions for the dust in $\lambda$-Orionis to argue that the PAH mass distribution correlates more strongly with AME than the total dust mass.

A third area of progress, and the one we focus on in this work, is improved maps of PAH emission. For AME physics, perhaps the most useful PAH feature is the 3.3 \micron\ feature that corresponds to C-H stretching modes.  This feature, being that of the highest energy, is emitted by the smallest PAHs \citep{Draine2021}, which are also most likely to be responsible for the AME if the spinning dust paradigm is correct \citep{Draine1998a,Ali-Haimoud2009}. 

The COBE/DIRBE satellite measured infrared emission over the entire sky in 10 bands spanning the infrared from 1.25 thorough 240 \micron. In the four shortest bands (1.25, 2.2, 3.5, and 4.9 \micron), only one (3.5 \micron) contains significant emission from PAH features (specifically, the 3.3 \micron\ feature). In this paper, we show that a simple template subtraction of the relevant DIRBE bands provides a good tracer of the most energetic PAH IR emission feature. We demonstrate this on the well-studied $\lambda$-Orionis region, which is a shell of neutral hydrogen and dust surrounding a Str\"{o}mgren sphere. Its associated HI column density ranges from $2$ to $50\times 10^{20}$ cm$^{-2}$ across the source \citep{Zhang1991}. Future work will extend this analysis to the full sky.    

In Section~\ref{sec:data}, we describe the COBE DIRBE and Planck data sets used.  In Section~\ref{sec:excess}, we describe our technique for isolating the 3.3 $\mu$m PAH feature in the DIRBE data. In Section~\ref{sec:PAH_signal}, we describe the construction of the PAH map used to test AME correlations.
Section~\ref{sec:correl} describes the correlations between the PAH signal and other Galactic signals (dust and AME) and their relative significance. In Section~\ref{sec:mask}, we quantify the effect of differences in masking on the relationships between the various correlations. We present a discussion in Section~\ref{sec:disc}. 

\section{Data} \label{sec:data}
\begin{figure*}
    \centering
    \includegraphics[width=8.5in]{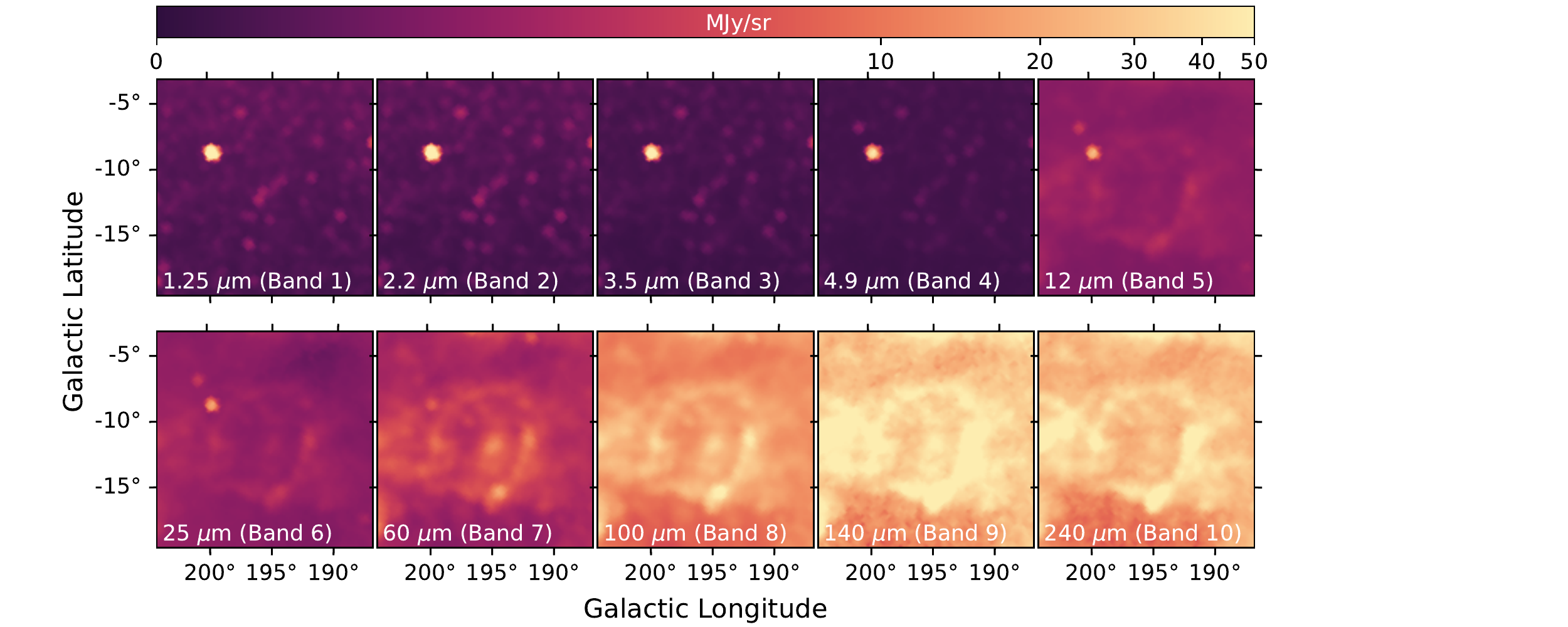}
    \caption{The $\lambda$-Orionis field as measured by all 10 of the DIRBE bands is shown. The ring structure is dominant in thermal dust emission at wavelengths $\gtrsim$12 \micron. At the four shortest wavelength bands, starlight dominates the intensity.}
    \label{fig:dirbe}
\end{figure*}

The COBE DIRBE experiment measured the sky in 10 infrared bands from 1.25 to 240 \micron\ with a $0.7^\circ\times0.7^\circ$ beam. The DIRBE experiment was calibrated to an internal reference, which makes it a good tool for all-sky measurements. In this paper, we use the Zodi-Subtracted Mission Average (ZSMA) Maps, which are hosted in their original format (quadrilateralized spherical cube projections in ecliptic coordinates) at LAMBDA\footnote{http://lambda.gsfc.nasa.gov}. For convenience in comparing with other all-sky data sets, we utilize HEALPix\footnote{http://healpix.sf.net} \citep{Gorski2005} formatted versions of these data sets\footnote{http://cade.irap.omp.eu/dokuwiki/doku.php?id=dirbe}. Maps of the $\lambda$-Orionis region are shown in Figure~\ref{fig:dirbe} for all 10 DIRBE channels. 

In addition to the DIRBE data, we utilize the Planck Commander AME map \citep{Planck_2015_X}. This map was derived from a spectral fit to Planck all-sky data using a two component spinning dust model.  Specifically, we utilize the amplitude of the variable-frequency component, given in units of $\mu$K$_{RJ}$, which is the dominant component in $\lambda$-Orionis \citep{Bell2019PASJ}. We verified that inclusion of the constant-frequency component does not affect our conclusions. The Planck AME map of $\lambda$-Orionis was originally discussed by \cite{Planck2016XXV}, in which the authors note the connection between the AME and the cold dust in the region.

\section{DIRBE Band 3 Excess Emission}\label{sec:excess}

In addition to the 3.5 \micron\ PAH feature, the DIRBE band 3 (3.5 \micron) image also contains starlight and low-level dust continuum emission. The adjacent band 2 (2.2 \micron) image also contains starlight and the continuum emission; however, there are no known PAH features in this band. Thus, we propose using band 2 as a template to isolate the PAH emission in band 3.

\begin{figure*}
    \centering
    \includegraphics[width=3.5in]{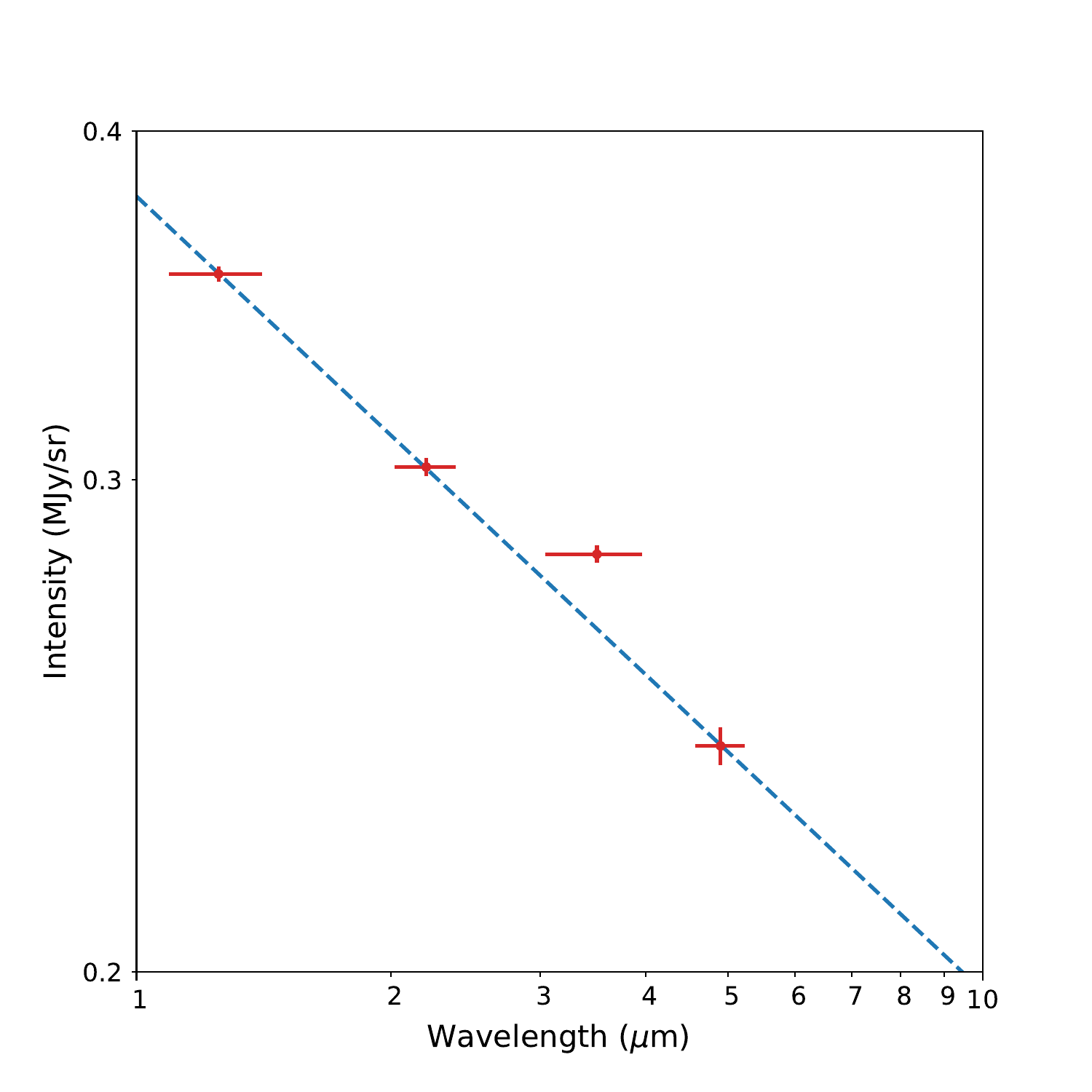}
    \includegraphics[width=3.25in]{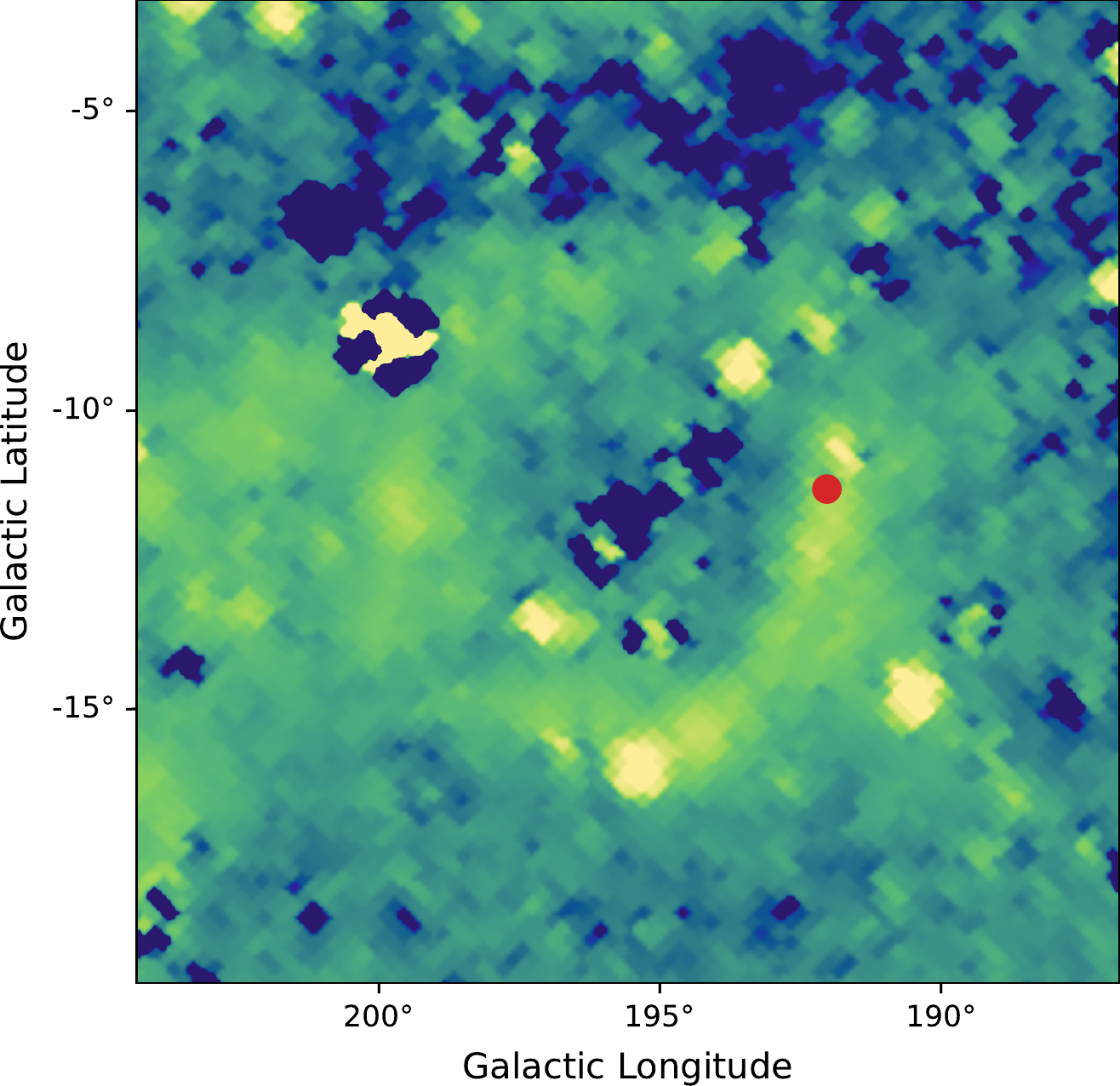}
    \caption{(left) The intensity of a position in the $\lambda$-Orionis field as measured by DIRBE bands 1--4 is shown. Horizontal error bars indicate the DIRBE bandwidths; vertical error bars are the DIRBE measurement uncertainties. The dashed line is a simple power-law fit to bands 1, 2, and 4. The excess emission above this power-law fit is visible in band 3.  (right) A preliminary PAH map of $\lambda$-Orionis is shown.  In this map, the continuum model for each pixel based on a power-law fit is subtracted from the DIRBE band 3 image. The ring structure is clearly seen. The position of the spectrum to the left is shown as a red filled circle.}
    \label{fig:Pahspec}
\end{figure*}

To demonstrate the viability of this approach, for each pixel in our field of view, we fit a power law to DIRBE bands 1, 2, and 4. The model fit along with the DIRBE intensities for a position on the $\lambda$-Orionis ring are shown in Figure~\ref{fig:Pahspec}. The excess emission in band 3 is visible, which we attribute to the PAH 3.3 \micron\ feature at this position.

We then subtract the model fit in each pixel (evaluated at 3.5~$\mu$m) from the DIRBE band 3 signal to produce the image shown in Figure~\ref{fig:Pahspec} (right). The structure of the ring is clearly visible in this image; in the band 3 image alone, it is clear that the ring structure is subdominant to starlight (see Fig.~\ref{fig:dirbe}.) The remaining bright point-like sources in the map are likely residual stellar sources that dominate these pixels and are not completely removed by this technique. This is a potential source of systematics, which we attempt to quantify by varying the masking of our eventual PAH image in Section~\ref{sec:mask}.
 
Though this map provides a validation of excess emission in band 3 that is associated with the 3.3 \micron\ PAH feature, the presence of thermal dust emission in band 4 is likely to result in an underestimate of PAH emission using this technique. Therefore, in calculating the PAH map for testing the association between PAH emission and AME, we utilize an template subtraction approach that does not utilize band 4.

\section{PAH Signal Estimation}\label{sec:PAH_signal}
Our goal is to compare the correlation between PAH emission and AME with that between AME and FIR continuum emission from dust. Specifically, we would like to test whether the PAH emission shows a \emph{higher} correlation with AME than the FIR dust emission does.  As such, our approach produces an estimate of the PAH distribution with the maximum possible correlation with the Planck AME map within the constraints of DIRBE data. We then test it against dust--AME correlations over the same region. 

For our quantitative analysis, we adopt a template subtraction technique to find an estimate of the map of PAH emission, $I_{\mathrm{PAH}}$, as a linear combination of the DIRBE band 2 and band 3 maps ($I_{\text{2.2\micron}}$ and $I_{\text{3.5\micron}}$, respectively) 
\begin{equation}
    I_{\mathrm{PAH}} = I_{\text{3.5\micron}} - \alpha I_{\text{2.2\micron}}+\beta.
    \label{eq:template}
\end{equation}

For our correlation study, we use a $16.7^\circ\times 16.7^\circ$ square region of the sky that is centered at $(\ell,b)=(195.7^\circ,-11.6^\circ)$. This captures the $\lambda$-Orionis ring along with some marginal surrounding area. 

The first step in the process is to mask the areas of the field where stars are dominant.  To do this, we set a threshold value in the 2.2 \micron\ map and mask areas above this value. The threshold value chosen for this map production is 0.6\,MJy\,sr$^{-1}$. The validity of this selection will be addressed in Section~\ref{sec:mask}. We then fit for the values of $\alpha$ and $\beta$ that maximize the Spearman correlation coefficient of the unmasked pixels of the PAH map (found with Equation~\ref{eq:template}) with the AME map. The fit utilizes the Nelder--Mead method for optimization as implemented in \texttt{scipy} \citep{Virtanen2020SciPy}.  

\begin{figure}[!h]
    \centering
    \includegraphics[width=0.45\textwidth]{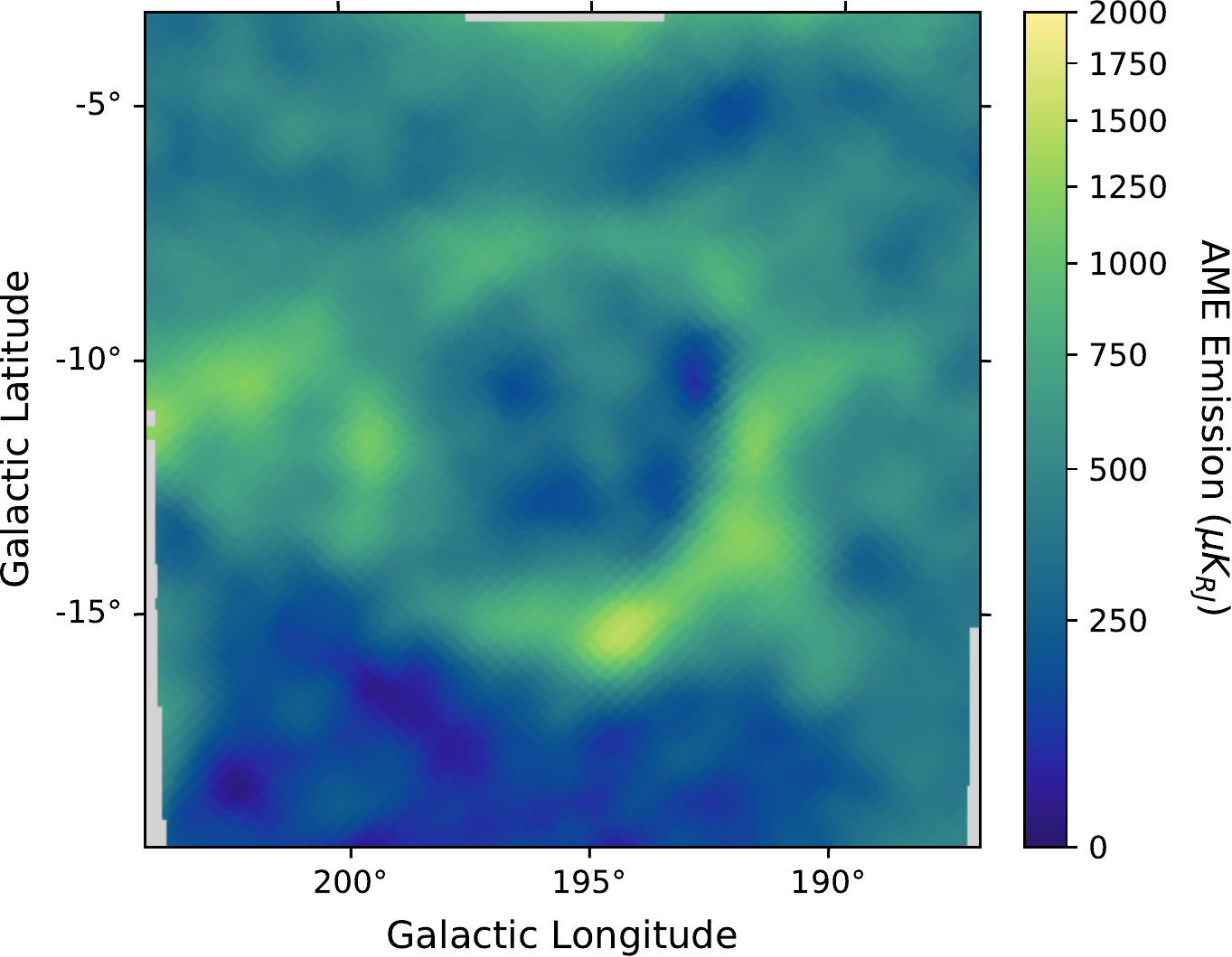}
    \includegraphics[width=0.45\textwidth]{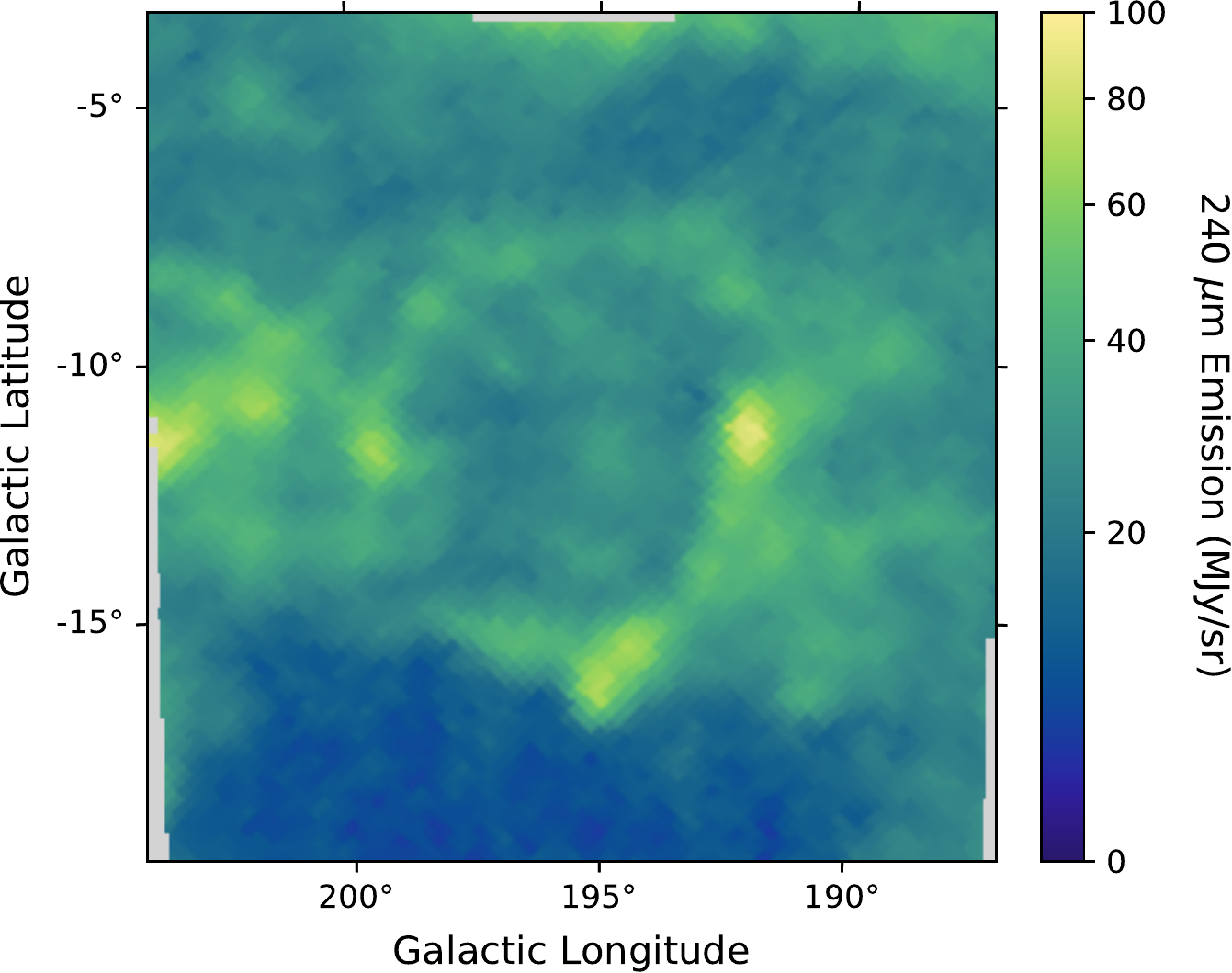}
    \includegraphics[width=0.45\textwidth]{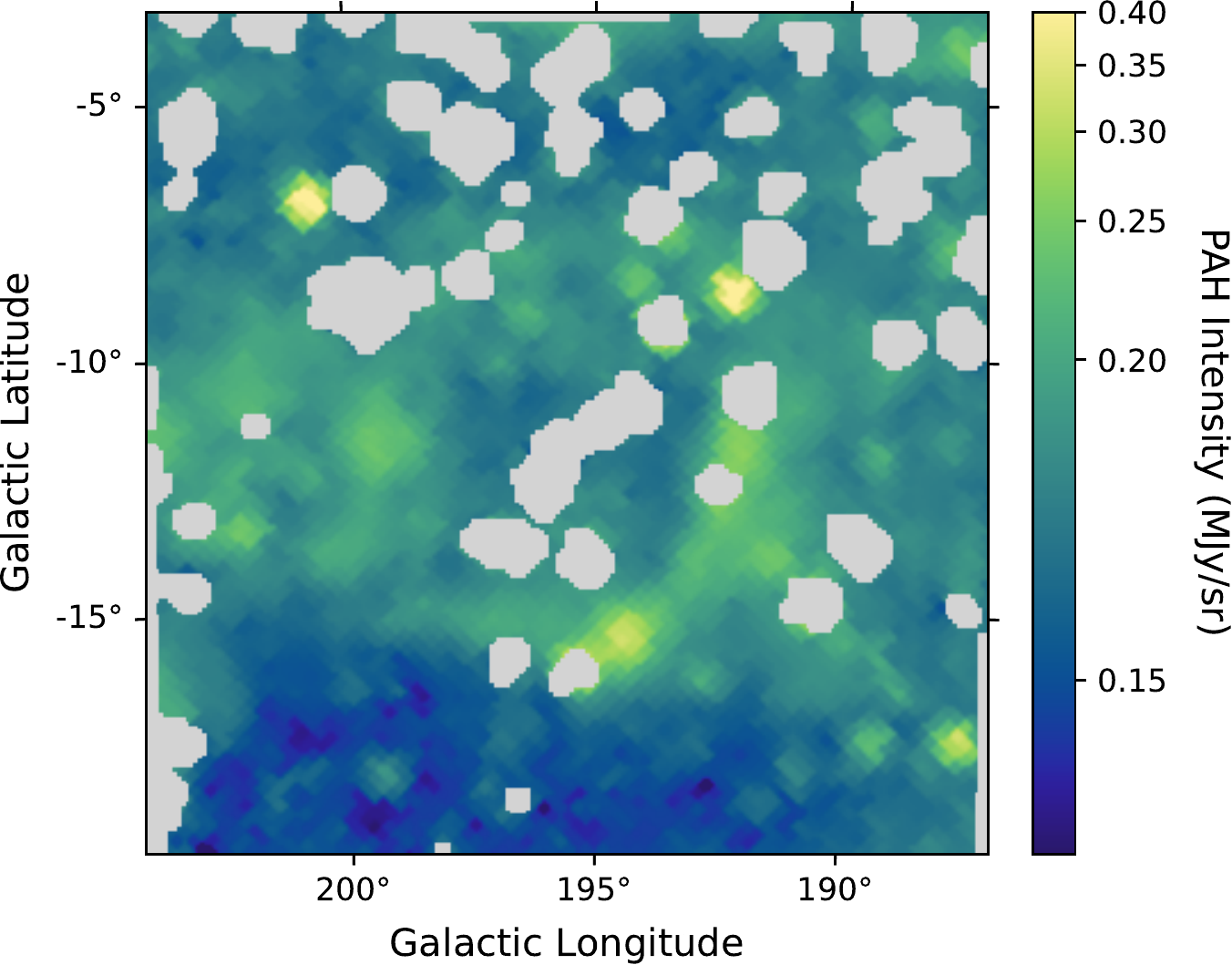}
    \caption{Top: the AME signal in the $\lambda$-Orionis region from the Planck Commander component separation is shown. Middle: the 240 \micron\ DIRBE image of the same region is representative of the thermal dust emission.} Bottom: the PAH map derived in this paper, which traces the 3.3 \micron\ C-H stretching modes. Here we use the mask corresponding to the 0.6\,MJy\,sr$^{-1}$ threshold for the 2.2 \micron\ map.
    \label{fig:PAH_signal}
\end{figure}

Using the resulting values $\alpha=0.49$ and $\beta=0.13$, we create the PAH map for the $\lambda$-Orionis region via Equation~\ref{eq:template}. It is apparent from Figure \ref{fig:PAH_signal} that ring-like structure is present in the $\lambda$-Orionis region, resembling the AME emission from Planck. The resulting PAH map with its final masking is shown in Figure~\ref{fig:PAH_signal} along with the AME map from Planck/Commander and the DIRBE/COBE 240 \micron\ dust map for the same region.

\section{Correlations}\label{sec:correl}
To test the possible relationship between the presence of PAH emission and AME, we analyze correlations between three data sets: our PAH map, the Planck Commander AME map, and the DIRBE 240 $\mu$m map that provides a tracer of thermal dust emission (see Fig.~\ref{fig:dirbe}: band 10).

\begin{figure}[!h]
    \centering
    \includegraphics[width=0.45\textwidth]{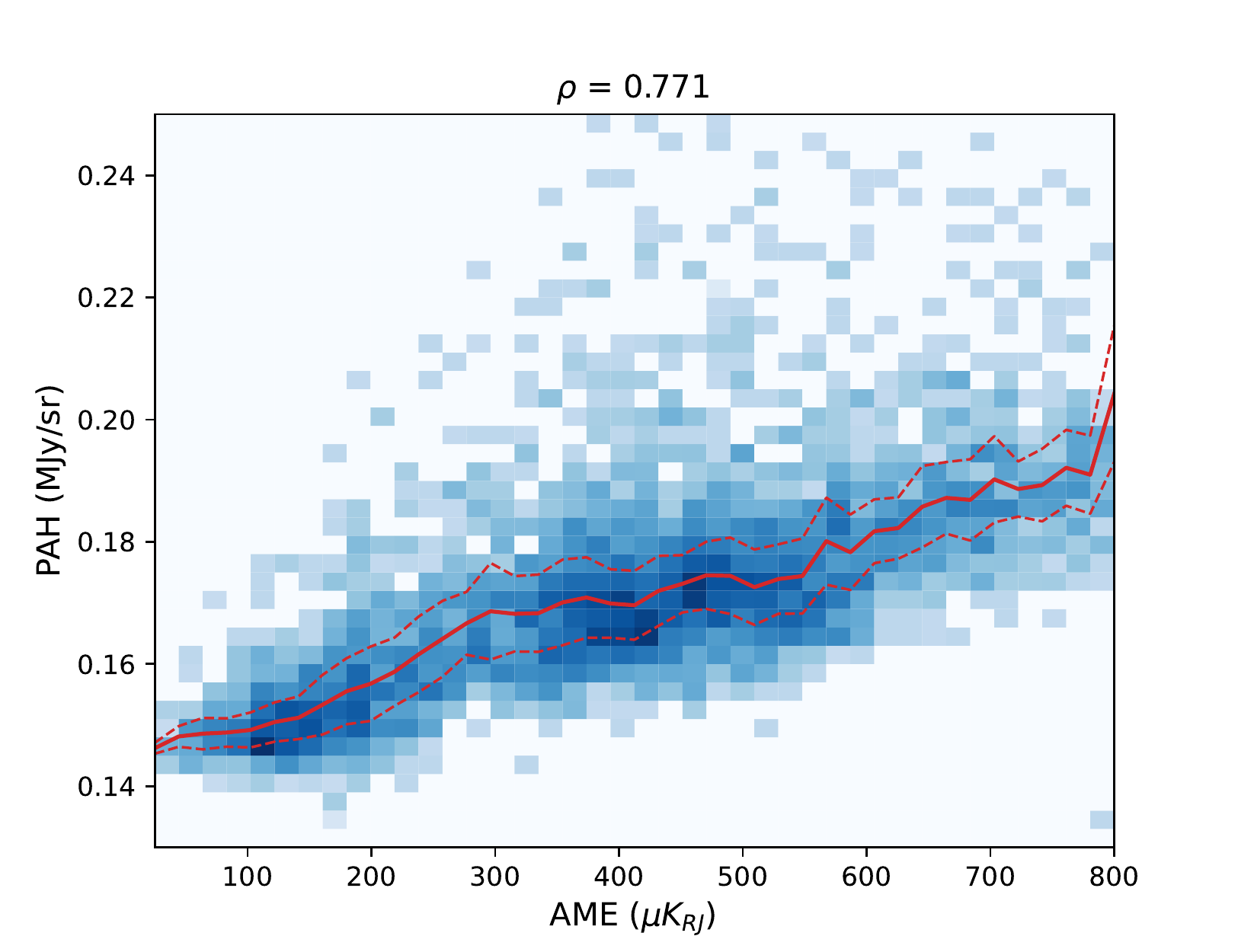}
    \includegraphics[width=0.45\textwidth]{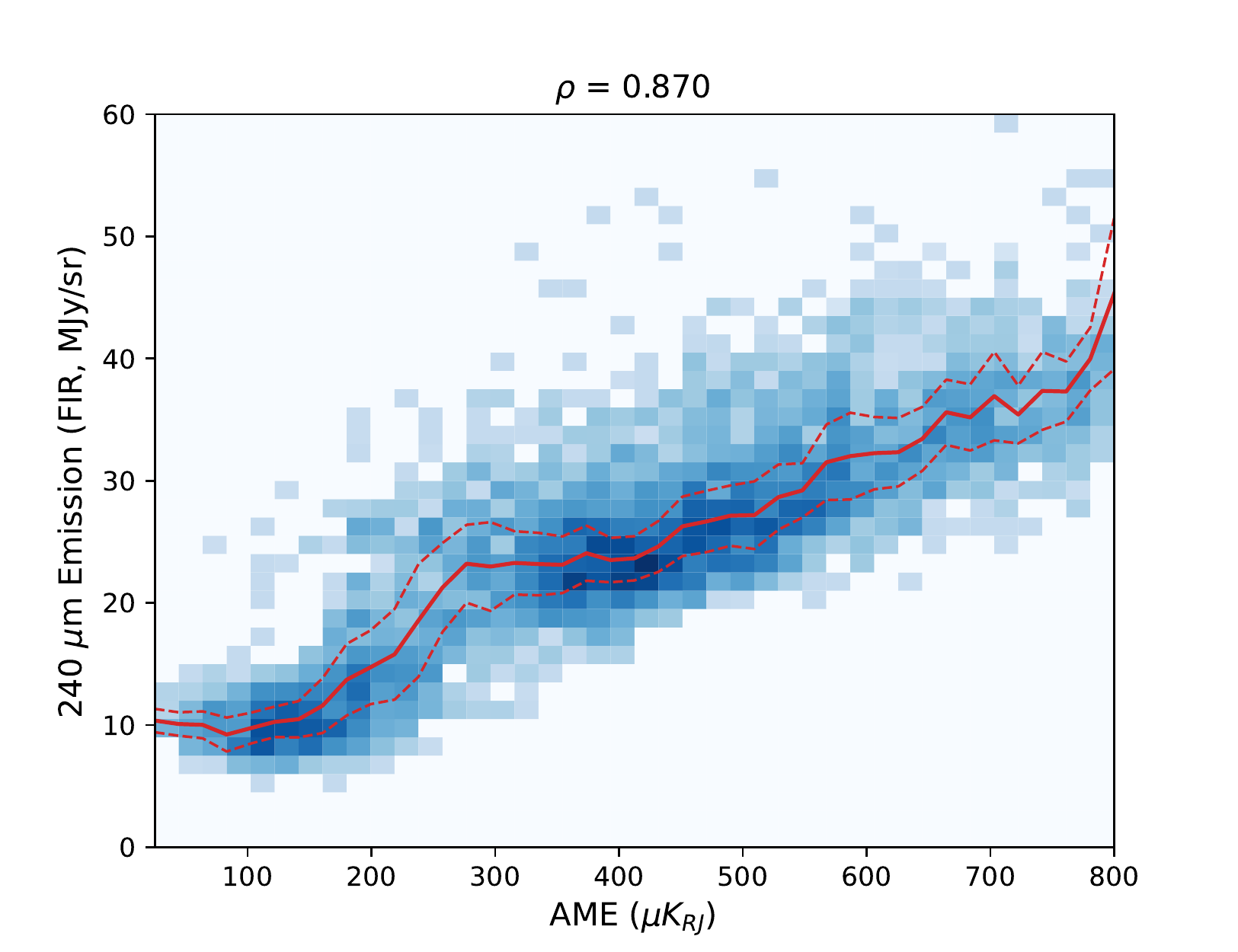}
    \includegraphics[width=0.45\textwidth]{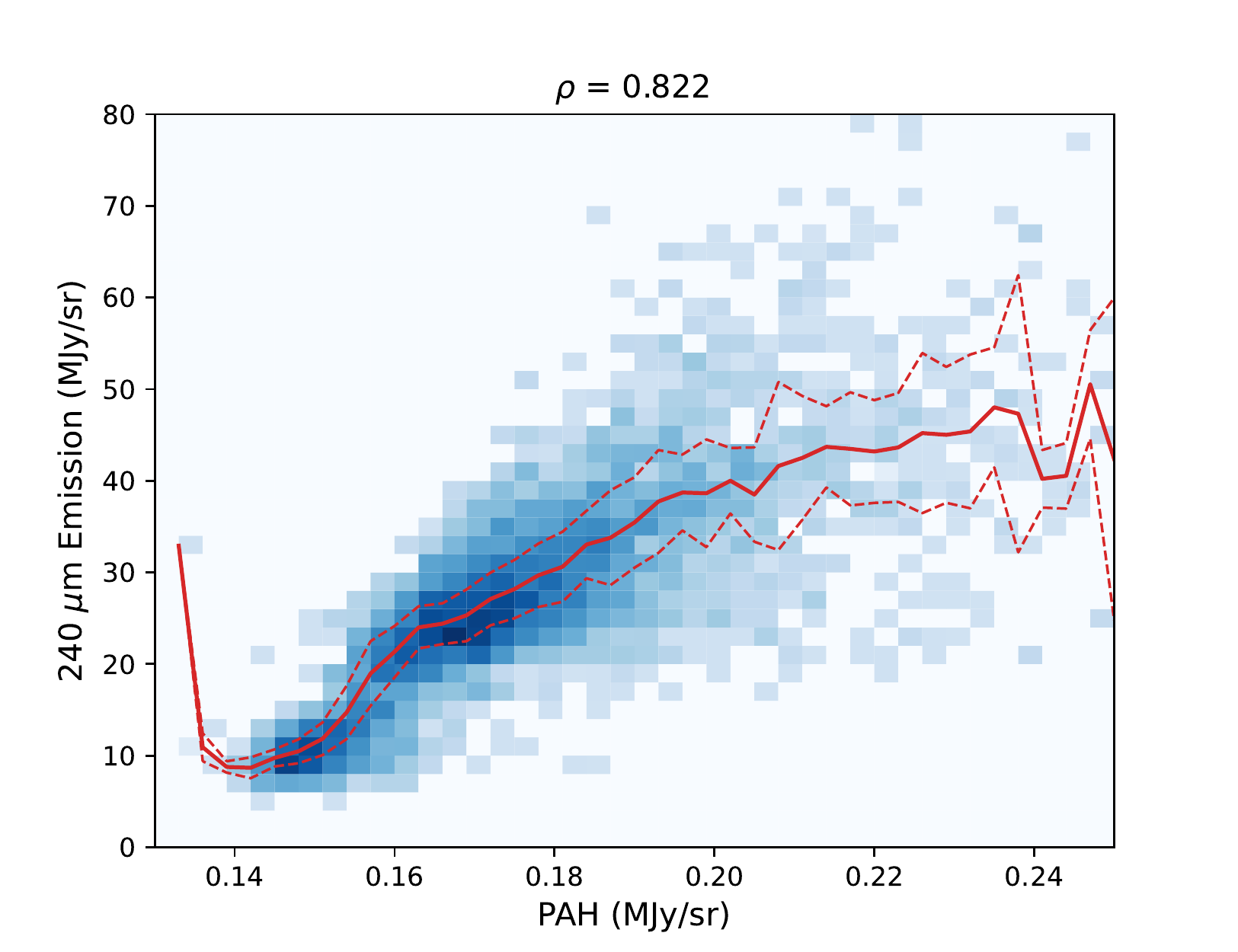}
    \caption{Correlations between maps are shown: (top) PAH emission vs. AME, (middle) FIR dust emission vs. AME, and (bottom) FIR dust emission vs. PAH emission.}
    \label{fig:PAH_AME}
\end{figure}

For each correlation, we plot a two-dimensional histogram of the map pixels, using 100 bins.  Results are shown in Figure~\ref{fig:PAH_AME}.  In each case, we plot the median in each bin as a solid red line, and the median average deviation is given by the dashed lines. 
We calculate the Spearman rank correlation coefficient, $\rho$, in each case. Results of these correlations are summarized in Table~\ref{tab:corr}.

To estimate the significance of the correlation coefficients in each case, we run simulations. For each pixel of the PAH, FIR (DIRBE 240 \micron), and AME maps, we redraw each data point from a Gaussian distribution that is characterized by a mean corresponding to the map pixel intensity value and a standard deviation equal to the uncertainty. We use the given DIRBE 240 \micron\ and AME Commander uncertainties; for the PAH uncertainty, we propagate the DIRBE bands 2 and 3 uncertainties ($\sigma_{\text{band 2}}$ and $\sigma_{\text{band 3}}$, respectively) as
\begin{equation}
    \sigma_{\text{PAH}}=\left(\sigma_{\text{band 3}}^2+\alpha^2\sigma_{\text{band 2}}^2\right)^{\frac{1}{2}}.
\end{equation}

\begin{figure}
    \centering
    \includegraphics[width=3.5in]{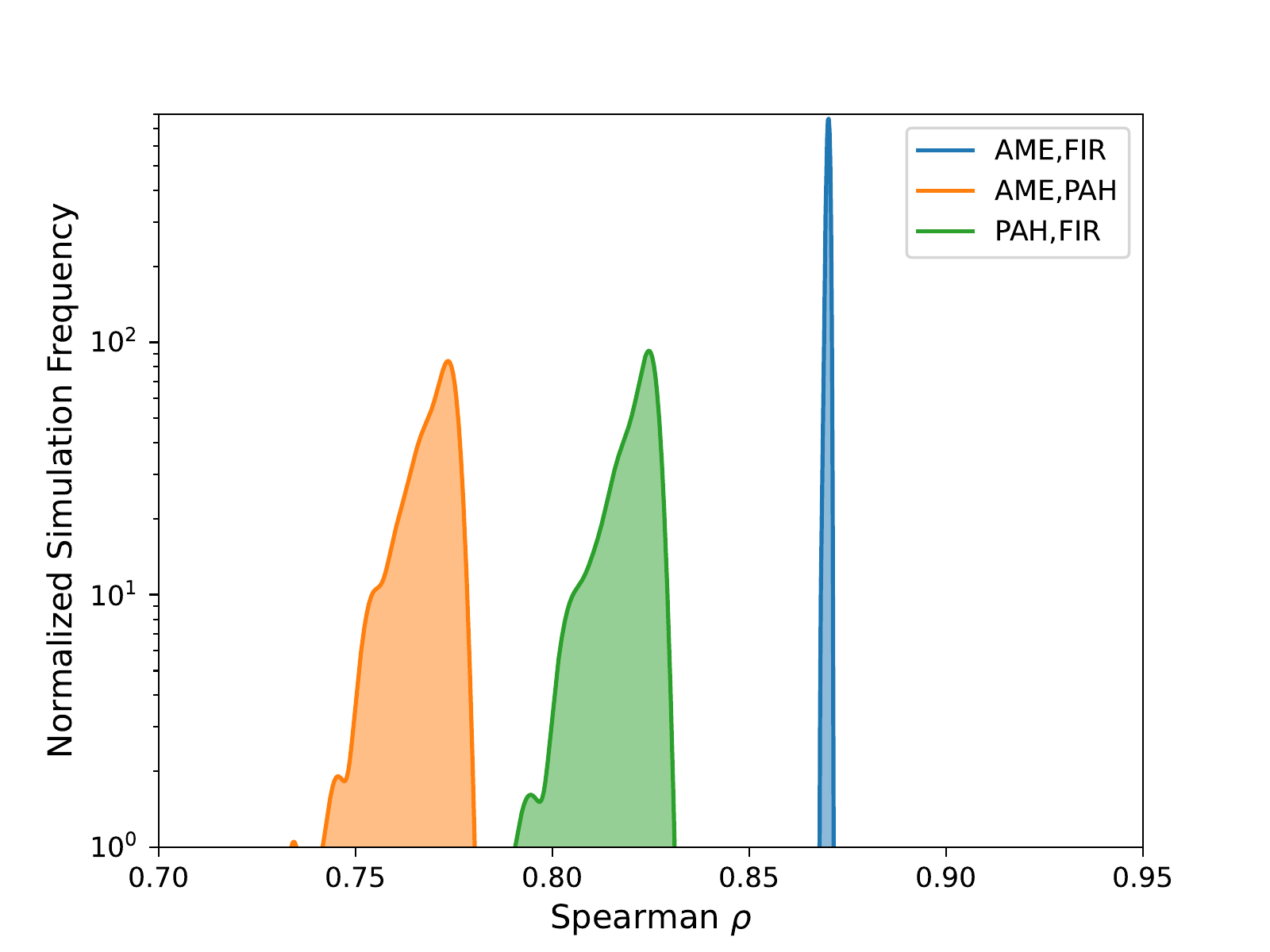}
    \caption{Histograms of Spearman rank coefficients calculated from 5000 simulations of the three data sets used in this work. This eliminates statistical uncertainty as the cause of the difference in correlation between AME--PAH and AME--Dust maps.}
    \label{fig:histsum}
\end{figure}

For each of these three maps, we simulate the maps 5000 times and calculate a set of Spearman rank coefficients. The distribution of each of these is shown in Figure~\ref{fig:histsum} and clearly indicates that the differences in correlation values are not a result of stochastic variation. The distributions of $\rho$ are asymmetric, as there are many more ways that random draws can decrease the correlation than increase it. Thus, the correlations we calculate from our data may be depressed from their actual values due to the noise by an amount that is similar to the width of this distribution. As a rough conservative estimate of the significance, we use the standard deviation of these distributions as an uncertainty on $\rho$ in each case. These have been tabulated in Table~\ref{tab:corr}. The uncertainties in Spearman $\rho$ for these comparisons are not sufficient to account for the differences between these correlations. Thus, we conclude that the PAH emission at 3.3 \micron\ is less correlated with AME than the FIR dust emission is if the noise in each map is statistical rather than systematic.

\begin{table}[]
    \centering
    \begin{tabular}{lll}\\
        \hline
        Data Set 1 & Data Set 2 & $\rho$\\\hline
        PAH & Commander AME  & $0.771\pm{0.01}$ \\
        240 $\mu$m DIRBE & Commander AME & $0.870\pm{0.001}$ \\
        240 $\mu$m DIRBE & PAH & $0.822\pm{0.01}$ \\\hline
    \end{tabular}
    \caption{Spearman Rank Correlation Coefficients}
    \label{tab:corr}
\end{table}

\section{Mask Dependence}\label{sec:mask}
In the above work, one of the potential systematic factors is the existence of starlight in the field. Emission from bright stars may not be completely subtracted in our estimate of the PAH map, and their residuals will result in an underestimate of the correlation between PAH and AME emission. To mitigate this, we have chosen a threshold value in our 2.2~\micron\ map, such that pixels with intensities above this threshold are ignored in the correlation calculations. However, our selection of this threshold value ($I_{2.2\,\mu\text{m}}=0.6$ MJy\,sr$^{-1}$) is somewhat arbitrary, and thus we are interested in estimating the sensitivity of our results to this parameter.

\begin{figure}[!h]
    \centering
    \includegraphics[width=0.45\textwidth]{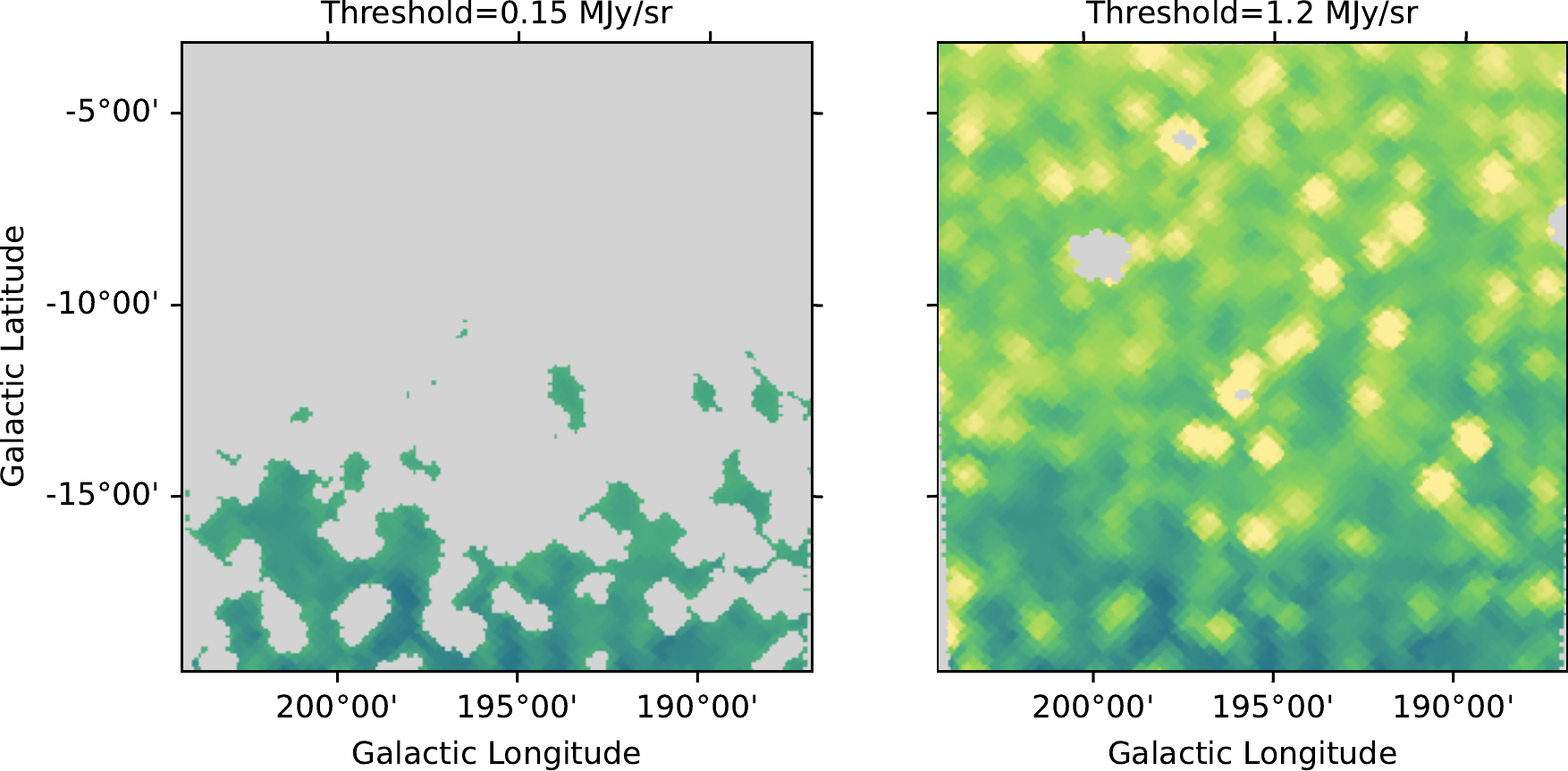}\\
    \includegraphics[width=0.45\textwidth]{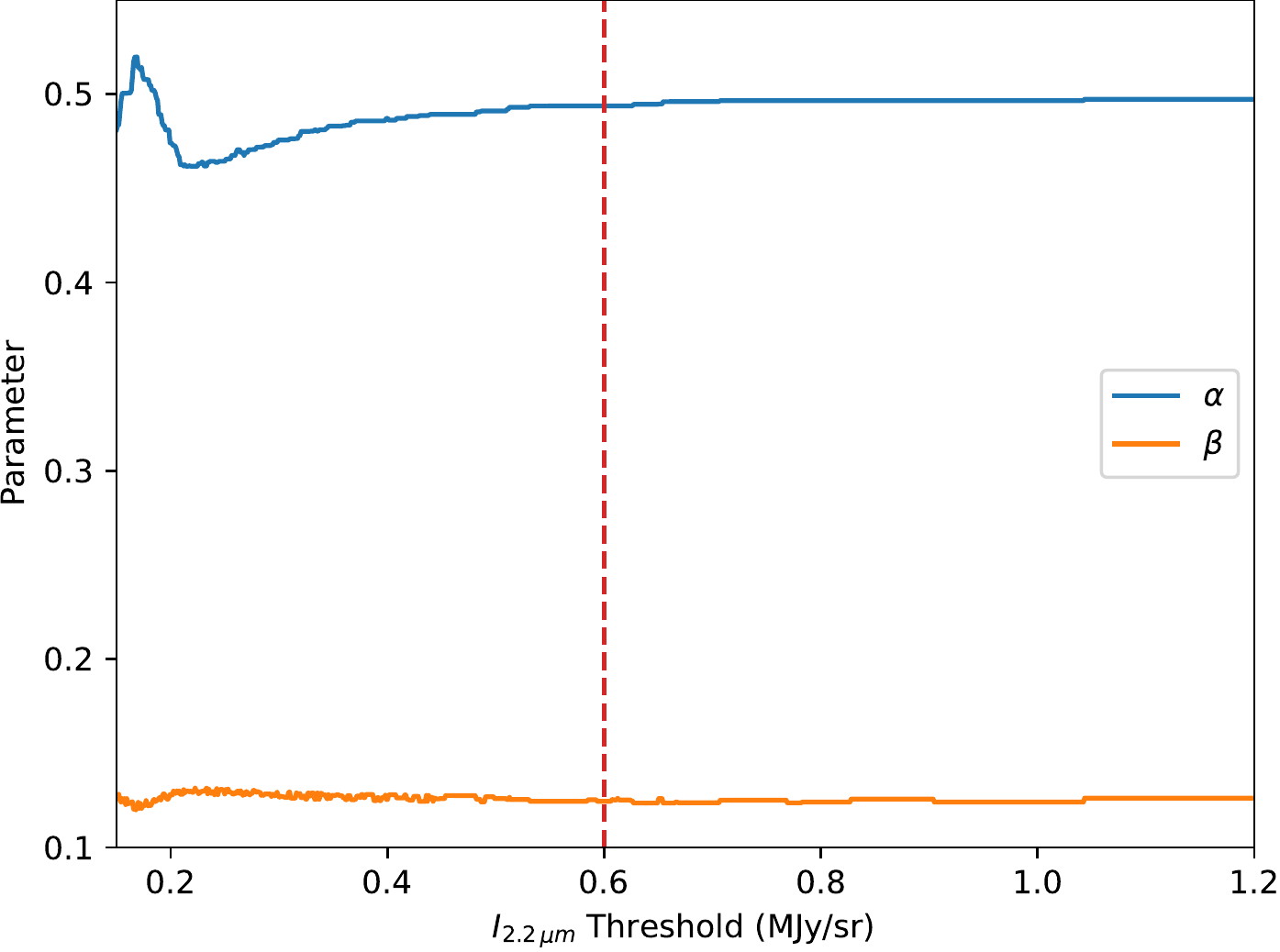}\\
    \includegraphics[width=0.45\textwidth]{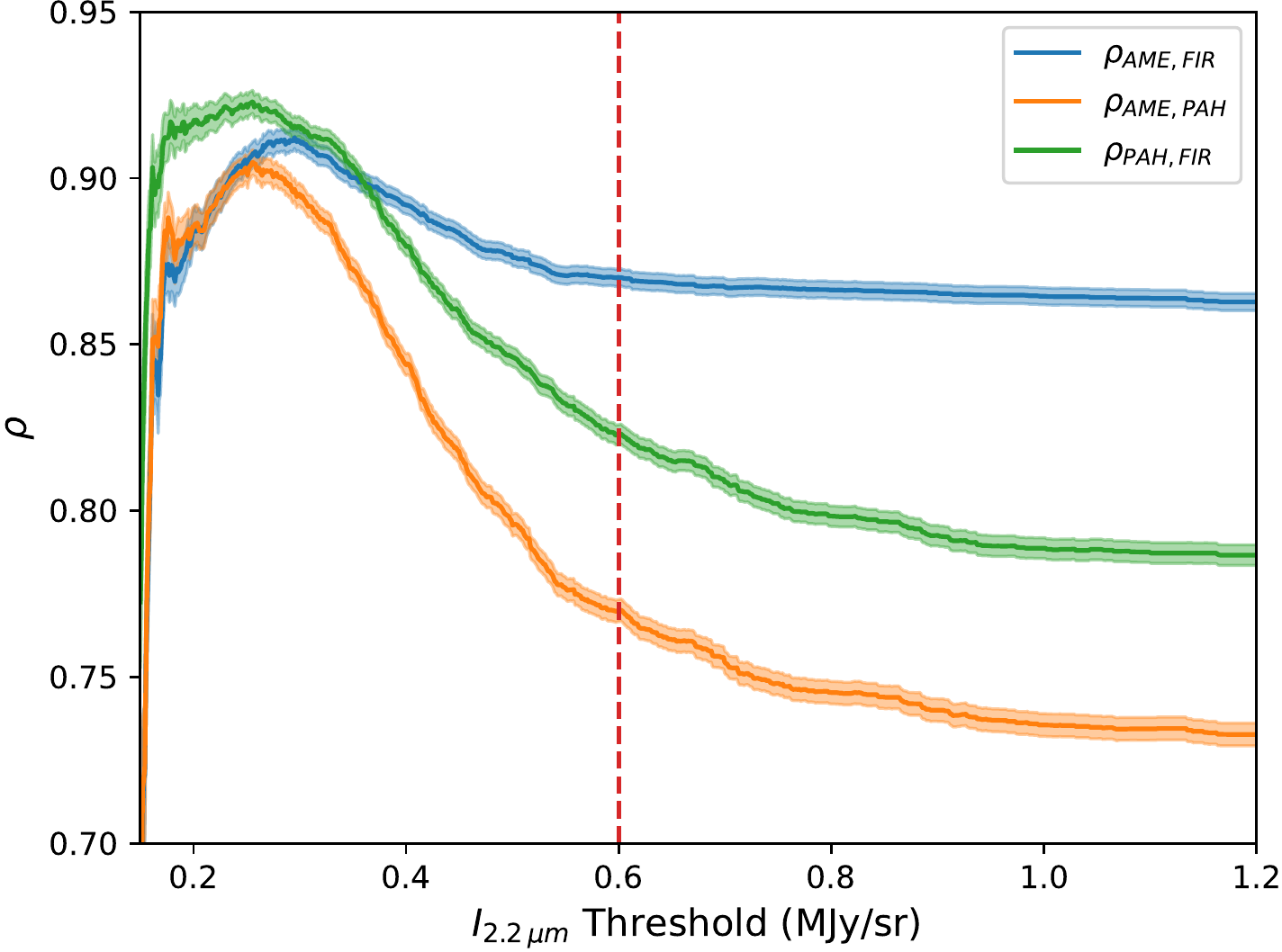}
    \caption{Top: masks using $I_{\text{2.2\micron}}$=0.15 MJy\,sr$^{-1}$ and $I_{\text{2.2\micron}}$=1.2 MJy\,sr$^{-1}$ are shown superposed on the $I_{\text{2.2\micron}}$ maps. These represent the masks for the extremes of the ranges of threshold values explored.  Center: the fit parameters for the PAH model are shown as a function of threshold value. Mean values are $\bar{\alpha}=0.49$ and $\bar{\beta}=0.13$. The vertical red dashed line indicates the threshold value of 0.6 MJy\,sr$^{-1}$ that was used for the results of the previous section for showing detailed maps and correlations. Bottom: Spearman correlation coefficients are shown as functions of the threshold value. The shaded band for each correlation indicates the ranges of possible values within the standard error of each (see the text for details.)  }
    \label{fig:corr_coeff}
\end{figure}

To quantify the dependence of the results of the paper on the the mask threshold value, we have tested the consistency of the fit parameters and correlations across for 1000 values of the threshold, linearly spaced between 0.15 and 1.2 MJy\,sr$^{-1}$. For each value, we calculate both the fit parameters ($\alpha$, $\beta$) and the various Spearman correlation coefficients to compare our PAH map with both AME and dust. The masks for the extremes of this range of threshold values are shown in Figure~\ref{fig:corr_coeff} (top), superposed on the $I_{\text{2.2\micron}}$ maps. 

The fit values as a function of the threshold value are shown in the center panel of Figure~\ref{fig:corr_coeff}. 
It can be seen that there is very little variation across the range of threshold. The mean values of the fit parameters are $\bar{\alpha}=0.49$ and $\bar{\beta}=0.13$.

Figure~\ref{fig:corr_coeff} (bottom) shows the Spearman correlation coefficient as a function of our threshold value. The shaded bands indicate the range of $\rho\pm\sigma$, where 
\begin{equation}
    \sigma\equiv\frac{\sqrt{1-\rho^2}}{\sqrt{n-2}},
\end{equation}
which is the expression for standard error of a correlation. Here $n$ is the number of pixels considered for each correlation.

Although a lower threshold produces modestly higher correlations,
the correlation between FIR dust emission and AME remains higher than the correlation between PAH emission and AME down to a mask threshold value of 0.3 MJy\,sr$^{-1}$. Below this, the curves are indistinguishable, and because of the relatively low number of pixels involved in the correlations at these low threshold values, no conclusions can be drawn. Future work that utilizes larger sky areas will be essential in understanding whether residual starlight is depressing the PAH--AME spatial correlation relative to that of FIR--AME.

\section{Discussion}\label{sec:disc}

There remains significant debate as to the origin of AME. The map derived from the residual 3.3 \micron\ PAH feature in the DIRBE 3.5 \micron\ band provides an additional means to trace PAHs, specifically those that are most likely to contribute to AME, assuming small spinning PAHs to be responsible for AME. The PAH emission map of $\lambda$-Orionis derived in this work correlates quite well with both FIR thermal dust emission and AME.  However, the correlation between the PAH and AME maps remains significantly below the correlation between AME and FIR thermal dust emission, even though our technique selects for the highest correlation possible between PAH emission and AME for this model. This result is in tension with the expectation that, if the smallest PAHs are the AME carriers, 3.3 \micron\ PAH emission and AME will be strongly positively correlated via their mutual correlation with PAH abundance. Alternative carriers of the AME, such as nano-silicates \citep{Hoang2016, Hensley2017}, provide a possible explanation for the observed lack of correlation.

However, the expectation of strong correlation between PAH emission and AME may not be correct in detail. \citet{Bell2019PASJ} fit dust SEDs in $\lambda$-Orionis using Hierarchical Bayesian techniques to estimate the correlation between the dust mass and AME to compare with that between PAH mass and AME. They find that PAH mass is better correlated with AME emission than total dust mass, and argue in support of electric dipole emission as the source of AME.  This is in agreement with \citet{Ysard2022} who compare dust/PAH correlations derived emission models with all-sky models across the full sky to show that nano-silicates are disfavored compared to a PAH-based origin of AME. 

Several authors have recently argued that environmental variations, which influence the amount, type, and excitation of PAH species present in a region of space, are critical to take into account in testing the relationship between the presence of PAH emission. For example, \citet{Hensley2022} utilized full-sky maps of $f_{\text{CNM}}$ to deduce that PAHs are systematically depleted in warm gas and that the emissivity per H atom of the PAH emission and AME may differ across physical environments, possibly explaining why mid-infrared PAH tracers have a weaker correlation than expected correlation with AME. Thus, even if the 3.3 \micron\ feature arises from the same population of PAHs that produce the AME, it is plausible that their correlation is reduced through these effects. 

This work provides a proof of concept of extracting the 3.3\,\micron\ PAH feature from DIRBE maps, and as such suggests future directions for further exploring this connection. Extending this technique over the entirety of the sky will enable more aggressive masking, which will aid in determining if residual starlight in our PAH map is responsible for depressing the AME--PAH correlation coefficient. The addition of 3.3 \micron\ spatial information across the sky could further inform models \citep[\emph{e.g.,}][]{Bell2019PASJ}, used to estimate correlations between the PAH mass and AME.  Finally, the 3.3 \micron\ map can be used in concert with other PAH emission maps (\emph{e.g.,} 12 \micron) to quantify how environmental conditions traced by the relative strengths of the PAH features affect the correlation between PAH emission and AME.

We thank the anonymous referee for helpful comments. This work was funded by NASA ADAP Grant NNX17AI91G to Villanova University.

\software{ \texttt{python, Ipython} \citep{Perez2007}, \texttt{numpy} \citep{vanderWalt2011}, \texttt{SciPy} \citep{Virtanen2020SciPy}, \texttt{matplotlib} \citep{Hunter2007}, \texttt{astropy} \citep{astropy:2013, astropy:2018}, \texttt{healpy} \citep{Gorski2005, healpy2019}}

\bibliography{PAH}{}
\bibliographystyle{aasjournal}

\end{document}